\documentstyle[aps,epsf,rotate,epsfig,preprint]{revtex}

\begin{document}

\draft

\title{Portfolio Optimization and the Random Magnet Problem} 

\author{Bernd Rosenow,$^{1,2}$ Vasiliki Plerou,$^{1,3}$ Parameswaran
Gopikrishnan,$^1$ \\ and H. Eugene Stanley$^1$}

\address{$^{1}$ Center for Polymer Studies and Department of Physics \\
Boston University, Boston, Massachusetts 02215, USA \\ $^{2}$ Institut
f\"ur Theoretische Physik, Universit\"at zu K\"oln, D--50937 K\"oln,Germany\\
$^{3}$ Department of Physics, Boston College, Chestnut Hill,
Massachusetts 02167, USA \\}

\date{\today}

\maketitle

\begin{abstract}

  Diversification of an investment into independently fluctuating
  assets reduces its risk. In reality, movement of assets are are
  mutually correlated and therefore knowledge of cross--correlations
  among asset price movements are of great importance.  Our results
  support the possibility that the problem of finding an investment in
  stocks which exposes invested funds to a minimum level of risk is
  analogous to the problem of finding the magnetization of a random
  magnet.  The interactions for this ``random magnet problem'' are
  given by the cross-correlation matrix {\bf \sf C} of stock returns.
  We find that random matrix theory allows us to make an estimate for
  {\bf \sf C} which outperforms the standard estimate in terms of
  constructing an investment which carries a minimum level of risk.

\end{abstract}


Challenging optimization problems are encountered in many branches of
science.  Typical examples include the traveling salesman problem
\cite{tsp,tsp2,tsp3} and the traveling tourist problem \cite{lima}.
Another type of optimization problem occurs when system parameters are
not accurately known and only estimates are available, such as in the
problem of finding the least risky investment in the stock market
which earns a given return.  Such an investment is called an optimal
portfolio. It has been suggested \cite{Galluccio98} that the
calculation of an optimal portfolio has an analogy in pure physics:
finding the ground state of a random magnet. However, the portfolio
optimization problem is more intricate due to the fact that 
many "system" parameters such as correlations are
not known with any degree of accuracy, but can only be estimated from
empirical data.

Two relevant pieces of information are necessary for an investor to
judge the quality of an investment: the investor must know (i) the
expected relative change in price (``return''), and (ii) the
uncertainty of the return (``risk''), usually measured by the standard
deviation of the returns over some preselected time intervals. Given
two investments with the same return, the investment with smaller risk
is preferred.  One way to reduce risk is to diversify the investment,
i.e., to buy stocks of not one, but of $N$ different companies
\cite{Elton95}. 
Diversifying the investment would work best if the fluctuations of
stock prices were completely uncorrelated; the risk would then
decrease with $N$ as $1/\sqrt{N}$. In reality, the price fluctuations
of different stocks are correlated.  The challenging optimization
problem is to choose the fraction of money to be invested into each
stock $m_i$ where $i$ runs over all $N$ stocks, in such a way as to
minimize the effect of correlations on risk of the N-stock portfolio.
We define  the return $G_i$ as the  relative price change of stock $i$,
$i=1...N$, and denote the expected total return by $R\equiv \sum_{i=1}^N m_i
R_i$ with $R_i=\langle G_i\rangle$, the return of an investment in
company $i$ over the investment period (in our empirical study half a
year).

The variance of $R$ is
\begin{equation}
D^2=\sum_{i,j=1}^N
(C_{ij} \sigma_i \sigma_j) m_i m_j\;, 
\label{eq.1}
\end{equation}
where the cross-correlation matrix {\bf \sf C} is the covariance
matrix normalized by the standard deviations $\{\sigma_i\}$ of
individual stocks \cite{Laloux99,Plerou99,GRPS00}.  To study the
influence of the cross-correlation matrix on investment decisions we
consider a straightforward investment problem first, where short
selling of stocks (i.e. borrowing stocks and selling them) is allowed
at no extra cost. In addition, we consider a problem where all the
capital is invested in stocks.  Enforcing the constraints of fixed
return $R$ and fixed total capital $\sum_{i=1}^N m_i=1$ by Lagrange
multipliers $\mu$ and $h$, the optimal portfolio is defined as the set
$\{m_i\}$ found by minimizing the function \cite{Elton95}
%
%
\begin{eqnarray}
  F ={1\over2}\sum_{i,j=1}^{N} \left( C_{ij}\, \sigma_i \sigma_j
  \right)\, m_i m_j - \mu \sum_{i=1}^{N} m_i R_i - h \sum_{i=1}^{N}
  m_i\;,
\label{free}
\end{eqnarray}
%
%
which is equivalent to the free energy of an Ising model with random
couplings $C_{ij}\, \sigma_i \sigma_j$ and a random magnetic field $
R_i$. From a physics point of view, selecting an optimal portfolio
amounts to calculating the mean field magnetizations $ m_i$ of this
random Ising model with the constraint of total magnetization one.  An
analytical solution exists since the free energy is quadratic.  The
expected return $R$ is a monotonically increasing function of the
standard deviation $D$. Thus, for accepting a large standard deviation
(risk) the investor is rewarded with a high expected return.

For the calculation of an optimal portfolio, one requires the $2N$
expectation values for future returns and standard deviations of stock
returns, and estimates for the $N(N-1)/2$ independent elements $C_{ij}$.
In practice, returns and standard deviations are estimated by combining
historical values with the judgement of analysts \cite{BlLi92}.  In
contrast, cross-correlations are estimated purely from historical time
series as analysts usually have expertise  in a specific industry
and therefore have difficulties evaluating cross-correlations between
different industries.

The problem of estimating cross-correlations is similar to knowing only
Monte Carlo time series for the dynamics of spins and estimating the
interactions between them from their correlations. In this physics
problem, interactions are stationary in time and one can in principle
calculate the exact correlation matrix by using infinitely long time
series. In the stock market problem, correlations may not be stationary,
and the use of long time series may not be possible. Estimating
correlations from short time series is plagued by considerable
statistical error.

Random matrix theory (RMT) allows one to separate noise and information
in {\bf \sf C} by comparing the statistical properties of {\bf \sf C} to
the properties of a random control {\bf \sf R} constructed from {\it
i.i.d.} time series \cite{wigner,guhr}. Agreement between {\bf \sf C} and {\bf
\sf R} is a signature of noise, whereas deviations indicate meaningful
information \cite{Laloux99,Plerou99,GRPS00,DGRS00,Noh00,LaCiPoBo00,%
BJNPZ01,GPRGAS01}.  Specifically, it was found that only the few
eigenvectors with eigenvalues larger than the upper edge $\lambda_+$ of
the random part of {\bf \sf C} contain information about groups of
correlated firms \cite{GRPS00} and are useful for the construction of
optimal portfolios \cite{GRPS00,LaCiPoBo00,GPRGAS01}. Here, we go
considerably beyond the analysis in previous approaches.  We (i) compare
portfolios constructed with RMT methods to those constructed under the
standard assumption that the only common influence on different stocks
is the whole market and (ii) systematically study whether portfolios
constructed with the RMT method have the lowest possible risk.

We diagonalize {\bf \sf C} and rank-order its eigenvalues $\lambda_k$
such that $\lambda_{k+1} > \lambda_k$. To filter from {\bf \sf C} the
effects of the random part, we calculate the upper edge $\lambda_+$ of
the random part of {\bf \sf C} and find that $\lambda_{989}$ is the
smallest eigenvalue larger than $\lambda_+$.  In order to keep only the
part of {\bf \sf C} which contains information about correlated groups
of companies, we construct a `filtered' diagonal matrix
$\Lambda^\prime$, whose elements are
\begin{equation}
\Lambda_{ii}'\equiv\cases{
        0 & $1\leq i<989$ \cr
\lambda_i & $989\leq i\leq 1000$.}
\label{lambda}
\end{equation}
We obtain the {\it filtered} correlation matrix {\bf \sf C$^\prime$} by
transforming $\Lambda^\prime$ to the basis of {\bf \sf C}.  In addition,
we set the diagonal elements to one as every time series is completely
correlated with itself.

We compare the proposed method to a method in which the
cross-correlation matrix {\bf \sf C}$^{\prime \prime}$ is calculated
under the assumption that the only common influence on two stocks is
the whole market, i.e. the one factor model \cite{Elton95}. This
assumption is wide spread as on the one hand it is known that the
price of a market index as the S\&P500 (comprising the 500 largest US
stocks) has big influence on the price of individual stocks. On the
other hand, there have been many attempts to identify further factors
influencing the price of groups of stocks but none of these models was
found to have larger predictive power than the simple assumption that
only the market index influences stock prices \cite{Elton95}.  If
$G^{\rm M}(t)$ denotes the return of the market index (we use the
S\&P500 index), then the return of stock $i$ is $G_i(t) = R_i +
\beta_i G^{\rm M}(t) + \epsilon_i(t)$, where $\epsilon_i(t)$ are
random variables describing the component of the return of stock $i$
which is both independent of the market and independent of all other
stocks, and $\beta_i$ describes the response of stock $i$ to a price
change of the market.  The cross-correlation matrix {\bf \sf
  C}$^{\prime \prime}$ has elements $ C^{\prime \prime}_{ij}= \beta_i
\beta_j \sigma_{\rm M}^2 / (\sigma_i \sigma_j)$, where the standard
deviations of $G^{\rm M}$ and $\epsilon_i$ are $\sigma_{\rm M}$ and
$\sigma_i$,

To compare the quality of the RMT forecast with that of the control,
we analyze 30-min returns of $N=1000$ largest US stocks for the year
1994 \cite{TAQ}. We partition the year 1994 into two six-month periods
A and B and use the first period to calculate the RMT forecast {\bf
  \sf C$^{\prime}$} and the one-factor model forecast {\bf \sf
  C$^{\prime \prime}$} for the empirical matrix {\bf \sf C$^{\rm B}$}
in the second period.  As can be seen from Eq.(\ref{free}) one needs
the future returns and standard deviations as an input in addition to
{\bf \sf C} in order to calculate a portfolio. In practice these
quantities are estimated by specialists \cite{BlLi92}.  We use instead
the returns and volatilities actually realized in the second period
\cite{Elton95,LaCiPoBo00}.  In this way, we probe only the effect of
randomness in the correlations coefficients and our results are not
influenced by uncertainties in returns and standard deviations.  With
this input we calculate optimal portfolios, i.e., the weights $\{
m_i\}$ of investment made into stock $i$ for {\bf \sf C}$^{\rm A}$,
{\bf \sf C$^{\prime}$}, and {\bf \sf C$^{\prime \prime}$}.  Given
these weights, we calculate the risk for a given value of return.

\begin{figure}[hbt]
\narrowtext 
\centerline{
\epsfig{file=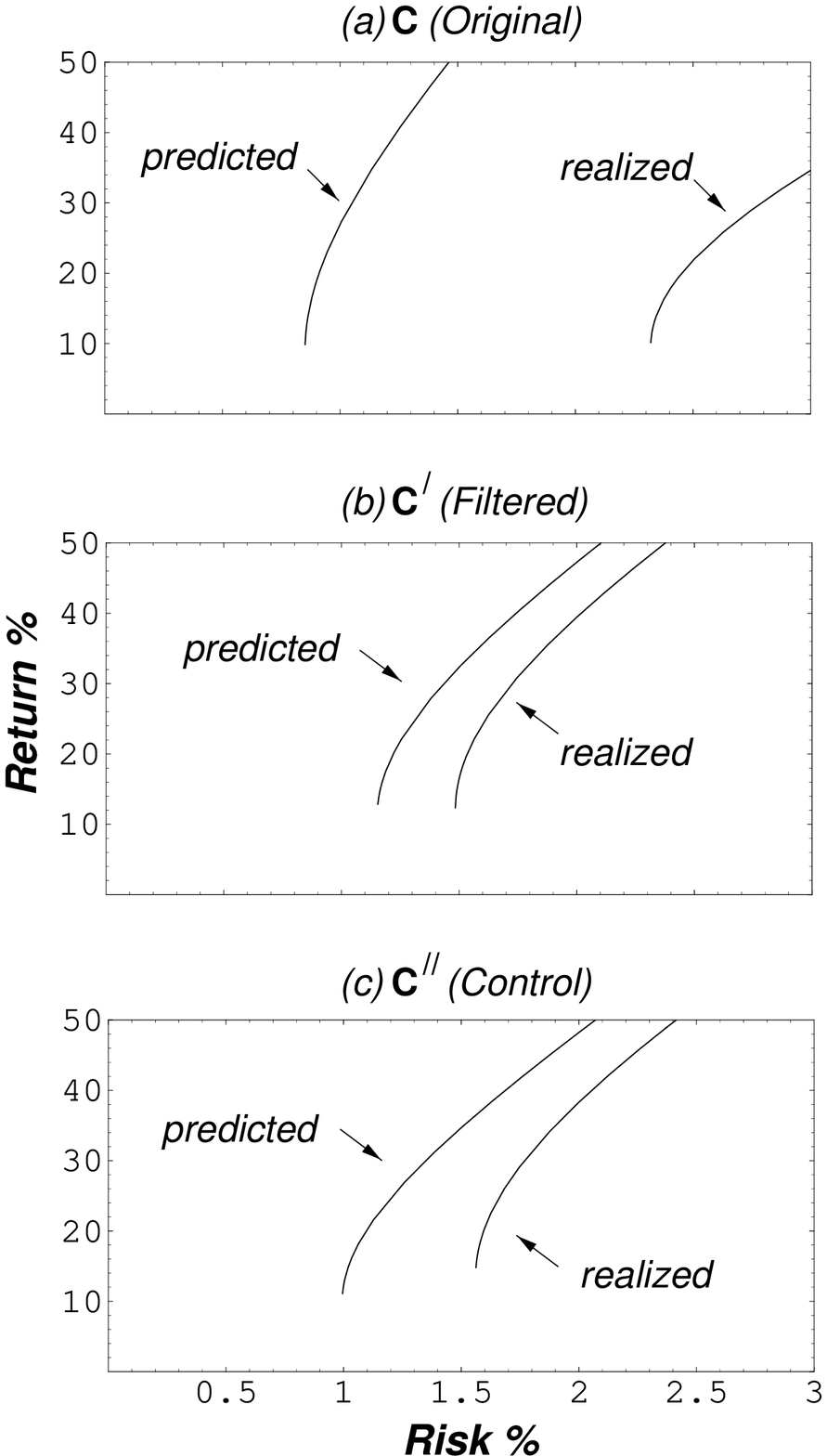,width=10cm}
}
\vspace{0.5cm}
\caption{Portfolio return $R$ as a function of risk $D$ for the
  families of optimal portfolios constructed from (a) the original
  matrix {\bf \sf C}, (b) the filtered matrix {\bf \sf C$^{\prime}$},
  and (c) the control {\bf \sf C$^{\prime \prime}$}.  The curves on the
    left show the predicted level of risk, whereas the curves on the
    right show the realized risk $D$ calculated using the correlation
    matrix {\bf \sf C$^{\rm B}$} for the second half of 1994. The ratio 
of realized to predicted risk is smallest for the RMT method (b), followed
by the control (c), and largest for the original matrix (a) }
\label{fig.2}
\end{figure}

We use three different tests to evaluate the performance of the  RMT method
as regards reducing risk.
First, we compare the {\it predicted\/} risk to the risk
which would have been realized if someone had invested using the
set of weights $\{m_i\}$.  We calculate this realized risk by using the
empirical cross-correlation matrix {\bf \sf C$^{\rm B}$} in
Eq.(\ref{eq.1}). In agreement with \cite{Elton95,LaCiPoBo00} we find
that the empirical matrix {\bf \sf C}$^{\rm A}$ is a very poor forecast
for {\bf \sf C}$^{\rm B}$ as the realized risk is 170\% higher than the
predicted one (relative difference).  For portfolios constructed with
the RMT forecast {\bf \sf C$^{\prime}$} \cite{LaCiPoBo00} and with the
standard forecast {\bf \sf C$^{\prime\prime}$} the relative difference
between predicted and realized risk is only 22\% and 33\%,
respectively. In addition to the higher accuracy in forecasting risk,
the realized risk for both {\bf \sf C$^{\prime}$} and {\bf \sf
C$^{\prime\prime}$} is considerably smaller than for the empirical
matrix {\bf \sf C}$^{\rm A}$ (Fig.~\ref{fig.2}).
 
\begin{figure}[hbt]
\narrowtext 
\centerline{
\epsfig{file=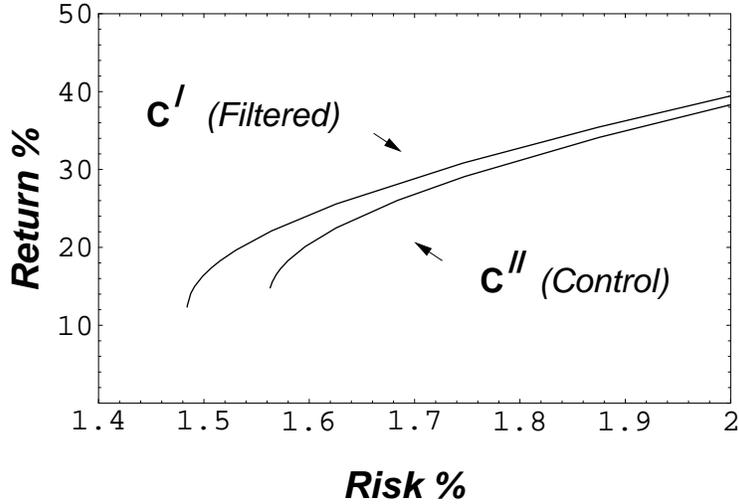,width=10cm}
}
\vspace{0.5cm}
\caption{ Comparison of the realized risk for the family of portfolios 
  constructed from {\bf \sf C$^{\prime}$} (RMT method) and {\bf \sf
    C$^{\prime \prime}$}(conventional method). For a given return, the
  RMT portfolios are characterized by a lower level of risk than the
the conventional portfolios. }
\label{fig.3}
\end{figure}

Next, we compare portfolios constructed with the standard forecast {\bf
\sf C$^{ \prime \prime}$} against portfolios constructed with the RMT
forecast {\bf \sf C$^{\prime}$}. We find that for a return of 15\% the
realized risk for the ``filtered'' portfolios is 5\% smaller than the
realized risk for the ``standard'' portfolios.  A similar reduction of
risk is also apparent for other expected returns (Fig.~\ref{fig.3}).  Thus, the
RMT method not only provides better estimates of future risks than the
standard method, but also allows to calculate portfolios with a 
considerably reduced realized risk.

\begin{figure}[hbt]
\narrowtext 
\centerline{
\epsfig{file=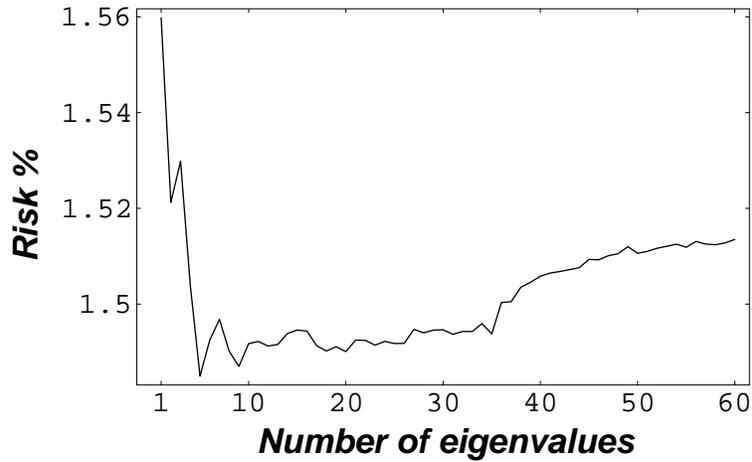,width=10cm}
}
\vspace{0.5cm}
\caption{Dependence of the realized risk on the number of eigenvalues kept 
  in the calculation of the cleaned cross-correlation matrix {\bf \sf
    C$^{\prime}$}. For this plot, the level of realized return is
  chosen as 15\% . RMT suggests that keeping 12 eigenvalues is
  the best choice for minimizing risk.}
\label{fig.4}
\end{figure}

Finally, we study whether the RMT method really suggests the optimal
number of eigenvalues which should be kept when constructing the
cleaned cross-correlation matrix. We calculate a family of
cross-correlation matrices {\bf \sf C$^{\prime}_p$} by keeping the
largest $p$ eigenvalues in the diagonal matrix {\bf \sf
$\Lambda^{\prime}$} instead of keeping 12 as in Eq.(\ref{lambda}). In
Fig.~\ref{fig.4} the realized risk for 15\% return is plotted against the number $p$ of
eigenvalues. For a range of $4 \leq p \leq 25$ the level of realized
risk fluctuates around the risk for $p=12$ (RMT suggestion) . Hence we
conclude that the RMT method provides a good estimate for the forecast of
future cross-correlations.

\begin{figure}[hbt]
\narrowtext 
\centerline{
\epsfig{file=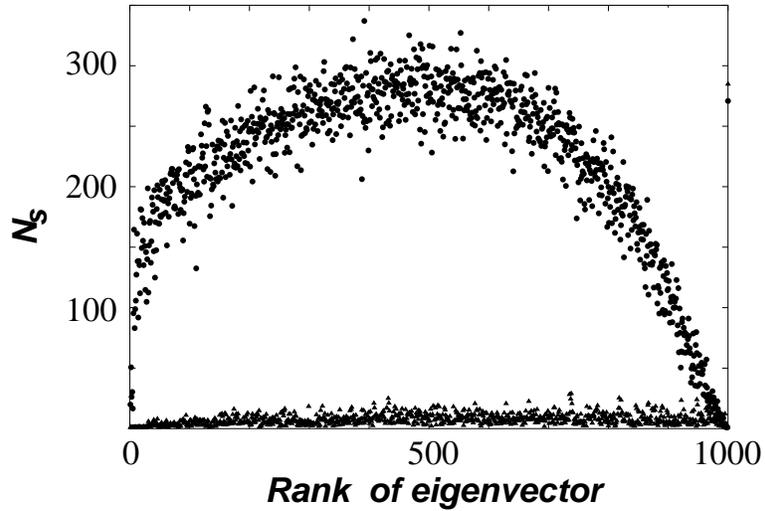,width=10cm}
}
\vspace{0.5cm}
\caption{The number $N_s$ of significant components of the eigenvectors of
  {\bf \sf C$^{\rm A}$} (circles) and {\bf \sf C$^{\prime}$}
  (triangles) is plotted against the rank of the eigenvector. $N_s$ is
  defined as one over the inverse participation ratio. Most of the
  eigenvectors of {\bf \sf C$^{\rm A}$} have a large $N_s$, whereas all
  but one of the eigenvectors of {\bf \sf C$^{\prime}$} have a small $N_s$
indicating individually fluctuating stocks or interactions between small
clusters of stocks. The last eigenvector with $N_s = 285$ 
describes the influence of the whole market and corresponds to a 
long range ferromagnetic interaction in the magnetic analogy.}
\label{fig.5}
\end{figure}

Having found that the cleaned cross-correlation matrix {\bf \sf
  C$^{\prime}$} is indeed a good choice for portfolio optimization, we
want to come back to the random magnet analogy and ask to what type of
random magnet the portfolio problem corresponds. For an investment
in the  stock market
as described by a
linear constraint fixing the total invested capital Eq.(2),
one cannot find a phase transition. Instead, the covariance
matrix acts like a susceptibility and the amount of invested capital
depends on the ratio of expected return to expected volatility of an
eigenmode. Alternatively, one can study  an investment in  futures
markets, where the investor is asked to leave a deposit
proportional to the value of the asset. This leads to a nonlinear
constraint $\sum_{i=1}^n |m_i|$ instead of the magnetic field term in
Eq.(2). Extrema of the free energy are described by coupled equations
\cite{Galluccio98} for the signs $S_i= {\rm sign}[ m_i]$
%
\begin{eqnarray}
S_i={\rm sign}\left[ \sum_{i=1}^N J_{ij}\left(\mu R_j + 
\nu S_j\right)\right]\ \ ,
\label{spin}
\end{eqnarray}
%
where $\left({J^{-1}}\right)_{ij}=C_{ij} \sigma_i \sigma_j$. In Ref.
\cite{Galluccio98} this optimization problem was studied for a
historical cross--correlation matrix  and found to be
related to spin glasses. Here, we argue that for the cleaned matrix
{\bf \sf C$^{\prime}$} one has to solve the problem of ferromagnetic
clusters in a random magnetic field. To see the difference, we compare
the eigenvectors of {\bf \sf C} and {\bf \sf C$^{\prime}$}. For each
eigenvector, we are interested in the number $N_s$ of significant
components which can be measured by one over the inverse participation
ratio (IPR) \cite{ipr}. We analyze the eigenvectors of the matrices
$C^{\rm A}_{ij}\sigma_i \sigma_j$ and $C^\prime_{ij}\sigma_i
\sigma_j$.  The number of significant components of the eigenvectors
of these matrices (which are also the eigenvectors of the inverse
matrices used in Eq.(\ref{spin})) is displayed in Fig.~\ref{fig.5}. Many of the
eigenvectors of {\bf \sf C}$^{\rm A}$ have more than 200 significant
components and describe long range frustrated interactions giving rise
to a spin glass type magnetic problem \cite{Galluccio98}.  On the
other hand, all but one of the eigenvectors of {\bf \sf C$^{\prime}$}
have less than 30 significant components. The eigenvector
corresponding to the largest eigenvalue has 285 significant components
and describes the influence of the whole market on the price dynamics
of an individual stock. In terms of the magnetic model, it describes a
long range ferromagnetic interaction.  The 999 eigenvectors with a
small number of significant component describe the fluctuations of
individual stocks or ferromagnetic interaction of small clusters of
stocks which can be identified as business sectors \cite{GRPS00}.
Hence we suggest that the magnetic problem equivalent to the portfolio problem
with a cleaned cross--correlation matrix is a random field
ferromagnet.

In summary, we used random matrix theory to estimate
cross-correlations and find that this method allows us to find
investments with substantially reduced risk compared to conventionally
used methods.  To accomplish this, we exploited a
formal analogy with the ``random magnet problem'', and analyzed the
cross-correlation matrix {\bf \sf C} of stock returns for short time
intervals extending over a one-year period.  We find an estimate for
{\bf \sf C} that outperforms the standard estimate, and allows us to
construct an investment which exposes the invested capital to only a
minimum level of risk.

{\it Acknowledgement\/}: We thank J.-P. Bouchaud, L.A.N. Amaral and
 L. Viceira for interesting discussions, and the National Science
 Foundation for support.

\end{document}